\documentclass[
a4paper, 
amsfonts, 
amssymb, 
amsmath, 
reprint, 
showkeys, 
nofootinbib, 
twoside,
superscriptaddress,
floatfix
]{revtex4-1}
\usepackage[english]{babel}
\usepackage[utf8]{inputenc}
\usepackage[colorinlistoftodos, color=green!40, prependcaption]{todonotes}
\usepackage{amsthm}
\usepackage{amsmath}
\usepackage{mathtools}
\usepackage{physics}
\usepackage{xcolor}
\usepackage[left=23mm,right=13mm,top=35mm,columnsep=15pt]{geometry} 
\usepackage{adjustbox}
\usepackage{placeins}
\usepackage[T1]{fontenc}
\usepackage{lipsum}
\usepackage{csquotes}

\usepackage{float}
\usepackage{graphicx}

\usepackage{fancyhdr}
\usepackage[pdftex, pdftitle={Article}, pdfauthor={Author}]{hyperref} 
\usepackage{color}

\definecolor{linkColor}{rgb}{0.0,0.7,0}
\hypersetup{colorlinks=true,linkcolor=linkColor, citecolor=linkColor, linktocpage}

\begin{document}

\title[Defect contrast with 4D-STEM]{Defect contrast with 4D-STEM: Understanding crystalline order with virtual detectors and beam modification}

\author{Stephanie M. Ribet}
\affiliation{Department of Materials Science and Engineering, Northwestern University, Evanston, IL, USA}
\affiliation{International Institute of Nanotechnology, Northwestern University, Evanston, IL, USA}
\affiliation{National Center for Electron Microscopy, Molecular Foundry, Lawrence Berkeley National Laboratory, Berkeley, CA, USA}

\author{Colin Ophus}
\affiliation{National Center for Electron Microscopy, Molecular Foundry, Lawrence Berkeley National Laboratory, Berkeley, CA, USA}

\author{Roberto dos Reis}
\email{roberto.reis@northwestern.edu}
\affiliation{Department of Materials Science and Engineering, Northwestern University, Evanston, IL, USA}
\affiliation{International Institute of Nanotechnology, Northwestern University, Evanston, IL, USA}
\affiliation{The NU\textit{ANCE} Center, Northwestern University, Evanston, IL, USA}

\author{Vinayak P. Dravid}
\email{v-dravid@northwestern.edu}
\affiliation{Department of Materials Science and Engineering, Northwestern University, Evanston, IL, USA}
\affiliation{International Institute of Nanotechnology, Northwestern University, Evanston, IL, USA}
\affiliation{The NU\textit{ANCE} Center, Northwestern University, Evanston, IL, USA}

\date{\today}
\begin{abstract}

Material properties strongly depend on the nature and concentration of defects. Characterizing these features may require nano- to atomic-scale resolution to establish structure-property relationships.  4D-STEM, a technique where diffraction patterns are acquired at a grid of points on the sample, provides a versatile method for highlighting defects.  Computational analysis of the diffraction patterns with virtual detectors produces images that can map material properties. Here, using multislice simulations, we explore different virtual detectors that can be applied to the diffraction patterns that go beyond the binary response functions that are possible using ordinary STEM detectors. Using graphene and lead titanate as model systems, we investigate the application of virtual detectors to study local order and in particular defects. We find that using a small convergence angle with a rotationally varying detector most efficiently highlights defect signals. With experimental graphene data, we demonstrate the effectiveness of these detectors in characterizing atomic features, including vacancies, as suggested in simulations. Phase and amplitude modification of the electron beam provides another process handle to change image contrast in a 4D-STEM experiment. We demonstrate how tailored electron beams can enhance signals from short-range order and how a vortex beam can be used to characterize local symmetry.

\end{abstract}
\maketitle

\maketitle

\section{Introduction} \label{sec:intro}

\begin{figure*}
\includegraphics[width=\linewidth]{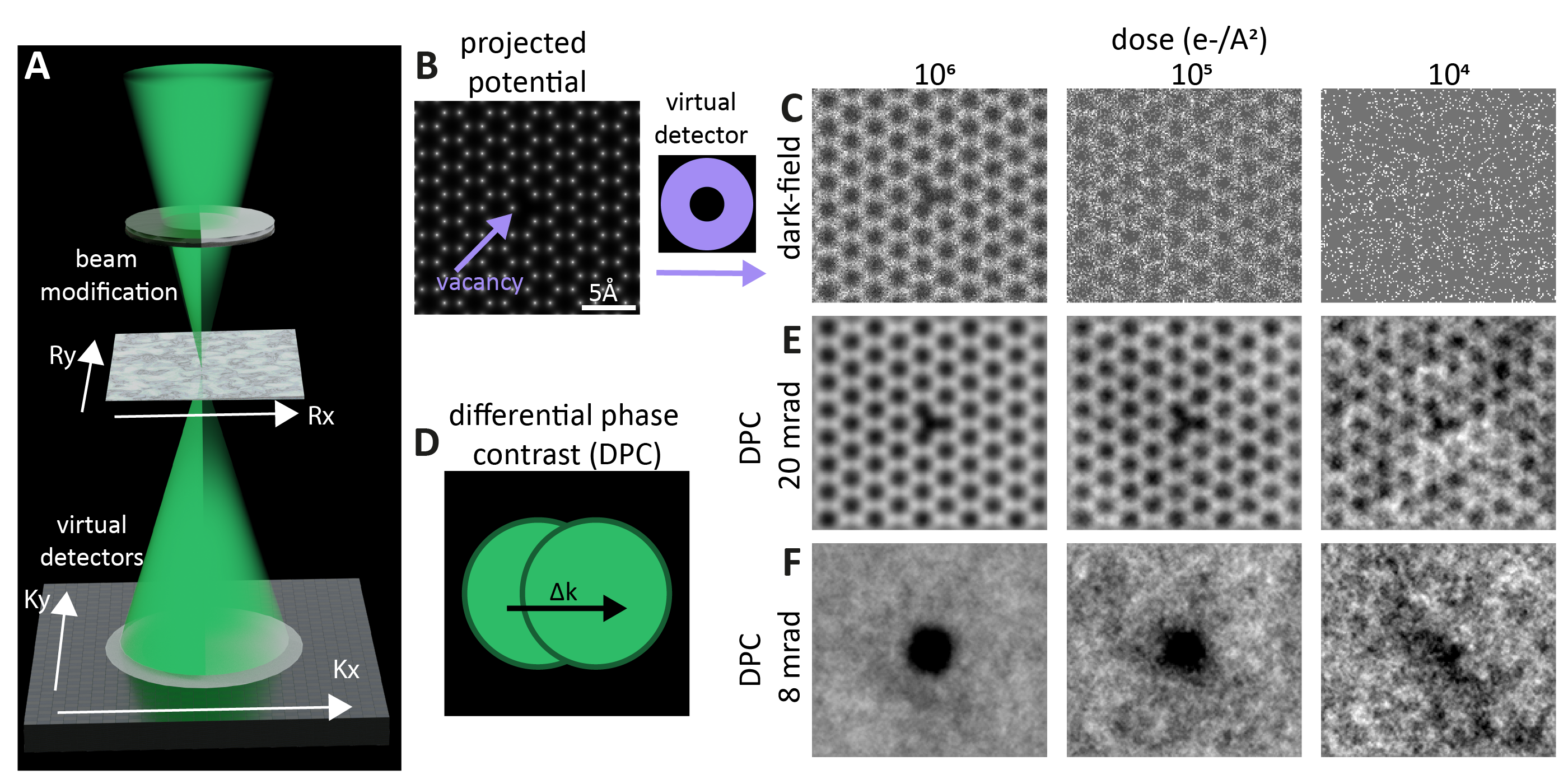}
\caption{(A) 4D-STEM schematic showing virtual detectors and beam modification. (B) Simulated potential of graphene with a vacancy. (C) Darkfield imaging is not dose efficient. (D) DPC can be used to recover the phase as shown in (E) and (F) A larger converge angle (E) provides atomic resolution contrast, while a smaller convergence angle (F) shows a stronger signal from the missing atom.}
\label{fig:overview}
\end{figure*}

The presence of defects disproportionately impacts the electronic and crystal structures of materials giving rise to many useful properties. As a consequence, the ability to identify and control the concentration and nature of defects has been explored for many years. For example, impurities in silicon change its electronic properties allowing for the development of solid-state devices~\citep{jacoboni1977review}. Defects at multiple length scales are known to enhance phonon scattering for more efficient heat conversion in thermoelectric systems~\citep{zheng2021defect}. Dislocations and grain boundaries play a key role in plastic deformation and resulting mechanical properties~\citep{van2006deformation}.  In nanocomposite materials, the topology and interfaces of dissimilar components strongly influence the structure-property relationships in these systems~\citep{distefano2020topology,ribet2021making}. Understanding defects and their impact on material properties is essential for technological development.

The presence of defects breaks long-range order in materials, which is reflected in local symmetry~\citep{zeng2011defect, sun2014anisotropic,liu2022visualizing}. There are a variety of ways to characterize order ranging from bulk scale property measurements capturing ensemble behavior to atomic-resolution imaging to identify individual structural changes~\citep{araujo2012defects}.  It is often important to reach nano- to atomic-scale resolution to establish form-function relationships, and the small probe size of scanning transmission electron microscopy (S/TEM) makes it a useful tool to study these features. Improvements in aberration correction of electron microscopes have led to the wide availability of sub-angstrom probe sizes that can be used for atomic resolution defect detection in many materials~\citep{dahmen2009background, krivanek2010atom}. 

In addition to traditional real space imaging approaches, local structural order can be probed by recording electron diffraction patterns.  A converged beam electron diffraction (CBED) pattern recorded in STEM reflects structural information about the area illuminated by the electron beam. In 4D-STEM or scanning diffraction experiments, a CBED pattern is recorded at each position defined in real space (Fig.\ref{fig:overview}A). These experiments have more recently been propelled forward thanks to the wide introduction of fast, pixelated direct electron detectors. One benefit of a 4D-STEM approach is the remarkable versatility in image reconstruction. For both experimental and simulated data, any mask can be applied as a virtual detector in post-processing or even during the experiment~\citep{plotkin2022100}, whereas conventional imaging approaches rely on binary detectors of fixed geometry. By collecting these large datasets and processing with virtual detectors, 4D-STEM can map material structure and properties not available in conventional experiments~\citep{ophus2019four}. 

There are many methods for signal enhancement in electron microscopy images to help probe local defects. In annular dark-field (ADF) STEM images, the contrast is sensitive to atomic number and thickness ~\citep{krivanek2010atom}.  These scattering cross sections can be used in physics-based models to find atomic positions even for sparse data~\citep{fatermans2018single}. Dark-field-based atomic counting algorithms are also sensitive to thickness and can be used for 3D material characterization including recovery of defects~\citep{van2011three, arslan20213d}. 
Neural-network-based deep learning analytical methods can be applied to characterize and retrieve material structural information with improved reliability and throughput ~\citep{ziatdinov2017deep, madsen2018deep}. Denoising algorithms, such as finding low-rank representations of the data can improve signals from defects~\citep{zhang2020denoising}, especially in multidimensional datasets and in some cases also in individual images~\citep{spiegelberg2018local}. Finally, more complex algorithms for multidimensional datasets, such as rotationally invariant variational autoencoders, can be used for enhancing signals and finding defects in materials~\citep{oxley2021probing}.
\color{black}

Despite improvements in hardware, software, and algorithms, there are still significant challenges in direct defect detection, especially when characterizing soft and hybrid materials~\citep{chen2020imaging, bustillo20214d, ribet2021making}.  These systems tend to be beam sensitive and may not withstand higher electron doses that are needed to resolve atomic features~\citep{egerton2022spatial}. Even with the improvement in the dynamic range of direct electron detectors~\citep{tate2016high, philipp2022very, haas2021high, plotkin2022100}, it can be challenging to capture weakly scattered signals. Therefore, we seek approaches that efficiently highlight defective sites and related symmetry information, without the need for atomic resolution studies. Easy defect detection can be useful for rapid characterization of materials and for intelligently informing microscopy parameters in \textit{on-the-fly} experiments. Moreover, there can be ambiguity about whether defects are inherent to the sample or caused from beam-specimen interactions, making it attractive to detect defects at lower doses.

Graphene, a two-dimensional structure, has attracted great attention due to its remarkable physical and electronic properties~\citep{geim2007rise}. Structural defects strongly influence graphene's performance and significant efforts have been devoted to characterizing them at multiple length scales with both spectroscopic and imaging techniques~\citep{araujo2012defects}.  Graphene is an ideal model system for testing the impact of beam modification and virtual detectors on 4D-STEM experiments. The simple scattering inherent to its two-dimensional nature makes it more straightforward to understand the origin of contrast in images. Moreover, graphene is the quintessential example of a weak phase object, meaning that it does not impart a large phase shift on the initial probe. In order to achieve reasonable contrast, especially for atomic resolution imaging, a high electron dose needs to be applied. The knock-on threshold for graphene is 86 kV~\citep{smith2001electron}, so a number of S/TEM studies have been able to image this structure at atomic resolution at or below the damage threshold~\citep{dahmen2009background, huang2011grains, ophus2015large, ziatdinov2017deep, ishikawa2018direct, madsen2018deep,  oleary2021increasing}.  These studies can serve as a benchmark for our work, as we search for efficient defect detection schemes, which can be extended to less robust materials. 

\begin{figure*}
\includegraphics[width=\linewidth]{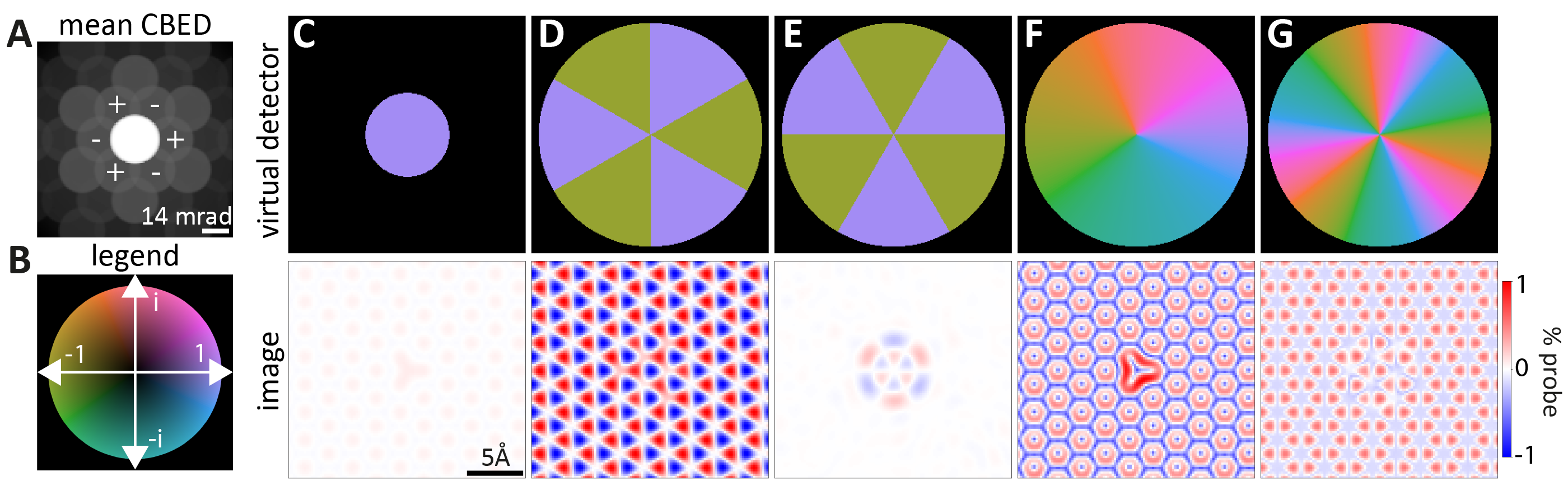}
\caption{(A) Graphene mean CBED pattern. (B) Legend for phase and amplitude of virtual detectors and electron beam. (C) Bright-field imaging does not efficiently capture the graphene signal. (D-E) Wedge detectors highlight coherent lattice or defects depending on orientation. (F) 1-fold complex detector shows the magnitude of CoM contrast. (G) The 3-fold complex detectors show atomic positions.}
\label{fig:detectors}
\end{figure*}

We first consider the application of virtual detectors to recover bright-field (BF) and ADF images in our model case of graphene with a vacancy (Fig.\ref{fig:overview}B). As seen in Fig.\ref{fig:overview}C, ADF imaging needs a relatively large number of electrons ($>$10${^4}e{^-}$\AA $^{-2}$) to yield an acceptable contrast to directly visualize the atomic positions.  BF imaging shows a lower dose efficiency in this case (Fig.\ref{fig:SI_BFDF}). Moreover, these images were formed with the most ideal experimental conditions, namely an aberration-free probe and a large convergence angle (20 mrad). The true resolution will be set by a balance between aberrations and the diffraction limitations of the convergence angle. Therefore, we seek a dose-efficient reconstruction method and a technique to highlight important signals derived from defective areas, even in resolution limited cases. 

For weakly scattering samples, phase contrast imaging approaches, such as differential phase contrast (DPC) or ptychography, lend themselves well to accurately recovering the object signal. The intensity distribution in the electron probe is shifted when it interacts with local variation in the sample potential, leading to a change in the center of mass of the transmitted electron beam (Fig.\ref{fig:overview}D). For a weak phase object, center-of-mass (CoM) images are directly related to the gradient of the sample electrostatic potential, and DPC images are formed by inverting the differential equation to plot the phase change of the electron probe~\citep{RN2,RN3, lazic2016phase, ishikawa2018direct, cao2018theory}. Figs.\ref{fig:overview}E-F show how much more efficiently DPC can capture structural information as compared to ADF with both a small and large electron probe. Ptychography refers to another family of reconstruction techniques that can be used to map the phase of the sample. Ptychography approaches can be used to solve for the probe and deconvolve it from the object to improve resolution and remove aberrations.  Despite the benefits of ptychography, this is a more computationally intensive method. Depending on the experiment and sample, either of these phase techniques may be the better approach~\citep{rodenburg2019ptychography, ophus2019four, oleary2021contrast}.

In this work, we explore how 4D-STEM can be used to study these interrelated issues of symmetry and defects. Electron microscopy simulations of beam-specimen interactions can be used in combination with experiments to guide data acquisition and analysis \citep{meyer2011experimental, jung2019unconventional, dacosta2021prismatic,parker2022scanning}.  These methods can also be used for rapid testing of new techniques before experimental realization \citep{kirkland1998advanced, madsen2021abtem}. Here we use multislice simulations to explore a wide range of virtual detectors in 4D-STEM experiments. In particular, we compare a variety of virtual detectors for detecting local order in two relevant materials, graphene and lead titanate (PbTiO$_3$). We test our design on experimental graphene data and show how these detectors can help highlight atomic contrast and defects. We study how beam modification can be used to amplify these defect features.  Moreover, the framework we develop for evaluating virtual detectors can be straightforwardly adapted to other challenges, in order to directly detect defects that cause a break in crystal lattice symmetry.

\section{Methods} \label{sec:methods}
\subsection*{Simulations}
Materials for simulations were built from structural files available through the Materials Project~\citep{Jain2013}. STEM simulations were performed using the \textit{ab}TEM~\citep{madsen2021abtem} multislice code based on methods laid out by Kirkland~\citep{kirkland1998advanced}. Image reconstructions were performed using custom virtual detectors and the py4DSTEM package~\citep{savitzky2021py4dstem}. Simulated datasets were created utilizing an acceleration voltage of 80 kV for graphene and 300 kV for PbTiO$_3$. Spherical aberrations, defocus, and tilt, were included in the construction of the STEM probe where noted. A perfect, aberration free probe is used as a ``control probe.'' Poisson noise was applied to simulate dose scaled in counts per probe at the end of multislice simulations. Because two-dimensional materials need a high number of frozen phonon configurations to converge~\citep{dacosta2021prismatic}, graphene simulations were run with 50 frozen phonons, while PbTiO$_3$ simulations were run with 10, and the standard deviation of the displacement was 0.05 \AA. For the 20 mrad convergence angle and 8 mrad convergence angle the collection angle of the bright field detector was $<$22 mrad and $<$9 mrad respectively. The ADF detector was 42 -100 mrad,with an inner collection angle matching \citep{ishikawa2018direct}.

\subsection*{Experimental data}
Experimental data was used from~\citep{oxley2021probing}. The data was acquired using a Nion UltraSTEM 100 operated at 60 kV with a convergence angle of 31 mrad.  Virtual detectors were applied for image reconstructions using py4DSTEM~\citep{savitzky2021py4dstem}. 
\color{black}

\subsection*{Virtual detector design}

In Fig.\ref{fig:detectors} we explore detectors, beyond conventional approaches.  Here we use both real and complex values in the detectors, which will be represented by the legend shown in Fig.\ref{fig:detectors}B. After applying a complex detector, we plot the magnitude of the sum of all pixels to form an image. Comparing images reconstructed with virtual detectors is a straightforward approach for evaluating 4D-STEM experiments -- in addition to the low computational cost and high speed of reconstructions, we can compare the signal to noise of images on a percent probe basis, which is a stand-in for electron dose~\citep{ophus2016efficient}. 

Analysis of 4D-STEM data is often driven by prior knowledge of the material structure or symmetry \citep{krajnak2020symmetry}. Based on the rotation symmetry of graphene, we would anticipate that 6-fold or 3-fold detectors would provide meaningful structural information. To evaluate the best detectors for analysis, we turned to principal component analysis (PCA). PCA has been used in other 4D-STEM experiments to guide virtual detector design ~\citep{han2018strain, nguyen2022angstrom}. Using a simulation of pristine graphene with a 20 mrad convergence angle we performed PCA analysis on the resulting data (Fig.\ref{fig:SI_PCA}). The first component shows the central beam. Components 2 and 3 are similar to x and y CoM virtual detectors, which underscores the effectiveness of these detectors for graphene. Component 4 shows 3-fold symmetry, which guided the virtual detector design in Fig.\ref{fig:detectors}.
\color{black}

\section{Results and Discussion} \label{sec:results}

\subsection*{Versatility of virtual detectors}

As described above, bright field imaging (Fig.\ref{fig:detectors}C) is not the most dose-efficient method. Fig.\ref{fig:detectors}D-E show wedge detectors that were designed with symmetry of graphene combined with the PCA approach. Based on Friedel's law~\citep{friedel1913symetries}, we expect opposite diffracted beams to be the same in magnitude and opposite in phase. Therefore, the intensity alternates in the areas of overlap between the first-order diffracted beams and the central disk. These different signals are highlighted by the `+' and `-' signs in Fig.\ref{fig:detectors}A. 

When the wedges are aligned with the lattice, as shown in Fig.\ref{fig:detectors}D, these signals interfere constructively, leading to strong intensities from atoms. Red and blue spots appear on atomic sites. The contrast from the vacancy is not easily identifiable with this detector due to the similar contrast of the missing atom.  When the detector is rotated 30$^\circ$, scattering from the lattice interfered destructively, so the areas corresponding to a perfect graphene lattice disappear.  However, the defect breaks long-range order and a strong signal from the missing atom arises. The combination of these detectors provides information about both order and sample orientation.

The magnitude of a 1-fold complex detector is analogous to the magnitude of a CoM reconstruction.  The image resulting from this detector (Fig.\ref{fig:detectors}F) shows a strong signal from both defect and atomic sites.  Higher-order rotationally varying detectors provide alternative information about local symmetry. In Fig.\ref{fig:detectors}G, we apply a 3-fold complex detector, which has easier interpretability than the 1-fold detector and strong signal from the atoms. The vacancy also has 3-fold symmetry, so it has similar contrast to the atoms in this imaging mode, although we can see small deviations in contrast around the missing atom.

\subsection*{Testing with experimental data}

\begin{figure*}
\includegraphics[width=\linewidth]{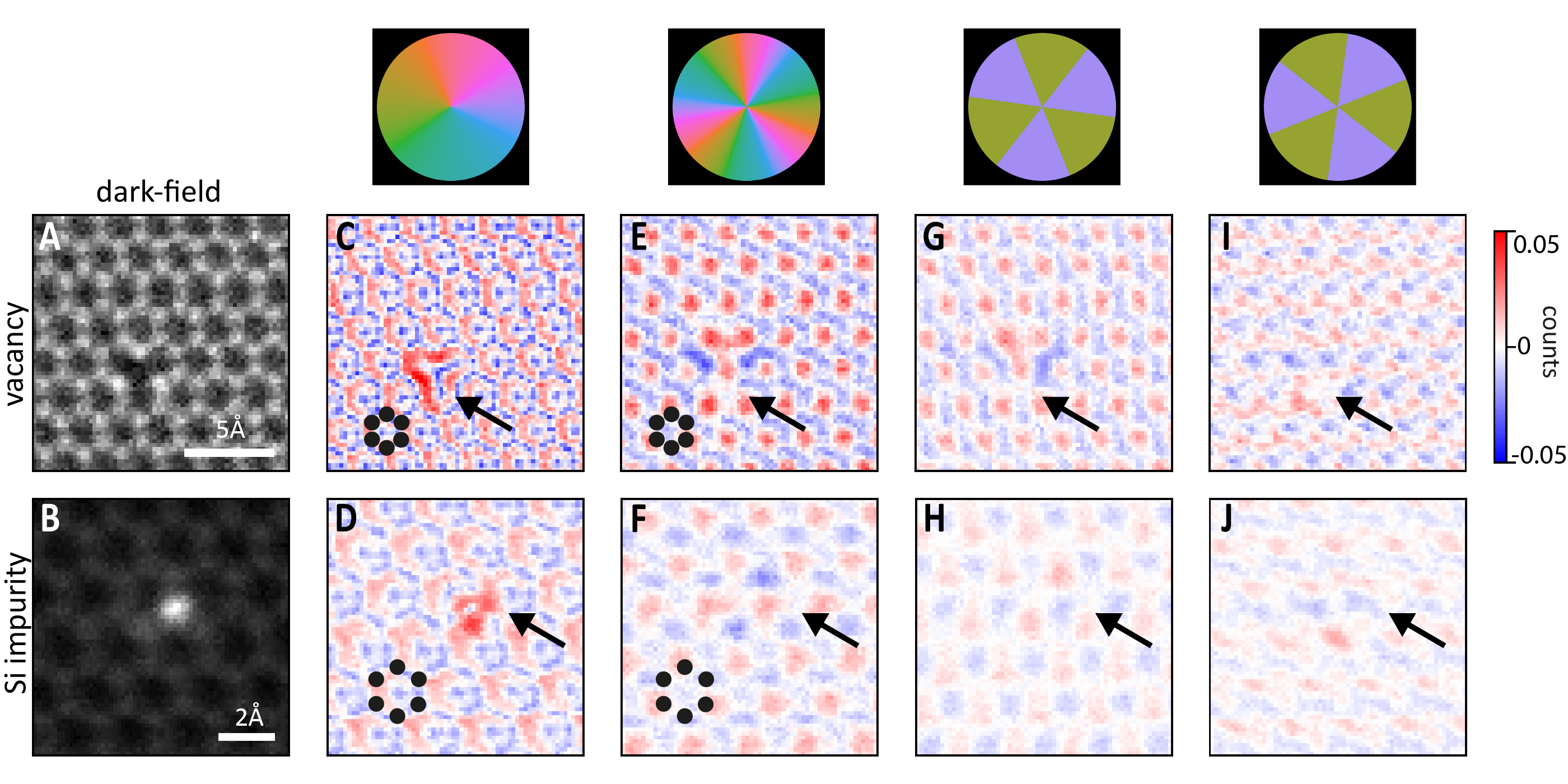}
\caption{Graphene with a vacancy and with Si impurity. (A-B) ADF for reference. (C-D) 1-fold complex detector show defect and graphene rings. (E-F) 3-fold complex detector shows graphene atomic positions more clearly than 1-fold detector. The vacancy appears as missing blue and red contrast. The Si impurity shows slightly higher atomic contrast than the graphene atoms in (F) but is not as clear as the in (D). (G-H) When aligned with the lattice, wedge detectors show atomic sites. (I) When the detectors are anti-aligned the vacancy appears as missing signal. (J) The Si impurity is highlighted.}
\label{fig:experimental}
\end{figure*}

In order to compare our simulations to experiments, we used 4D data from ~\cite{oxley2021probing} that studies defects in graphene. We used datasets with two different kinds of point defects, vacancies and Si substitutions. Fig.\ref{fig:experimental}A-B presents the simultaneously acquired ADF images, which we can use to compare against our reconstructions. Fig.\ref{fig:experimental}C-D show the images from the 1-fold complex detector. As with the simulated data, this matches the CoM magnitude images. The 3-fold complex detector (Fig.\ref{fig:experimental}E-F) provides a similar signal-to-noise value but more intuitive structural information.  The convergence angle used here provides atomic contrast, which is why the strong red and blue contrast from atoms is apparent. It is hard to see the Si impurity, as it is also in a 3-fold site with atomic contrast, highlighting a limitation of this detector. However, the vacancy is apparent, as we see missing signal at this position. 

As shown in Fig.\ref{fig:detectors}, images with the wedge detectors have strong rotational dependence. For a dataset with unknown relative rotation between real and reciprocal space, we can use the wedge detectors to solve for the rotation. The signal is maximized with a 8$^\circ$ rotation, and the resulting images (Fig.\ref{fig:experimental}G-H) are similar to the 3-fold complex detector reconstructions. The images in Fig.\ref{fig:detectors}I-J have weak contrast, which is expected given the destructive atomic interference. For the vacancy we see some signal missing from the defect, and we would expect the defect to be even further highlighted in an experiment with a smaller convergence angle. 
\color{black}

\subsection*{Understanding a large design space}

\begin{figure*}
\includegraphics[width=\linewidth]{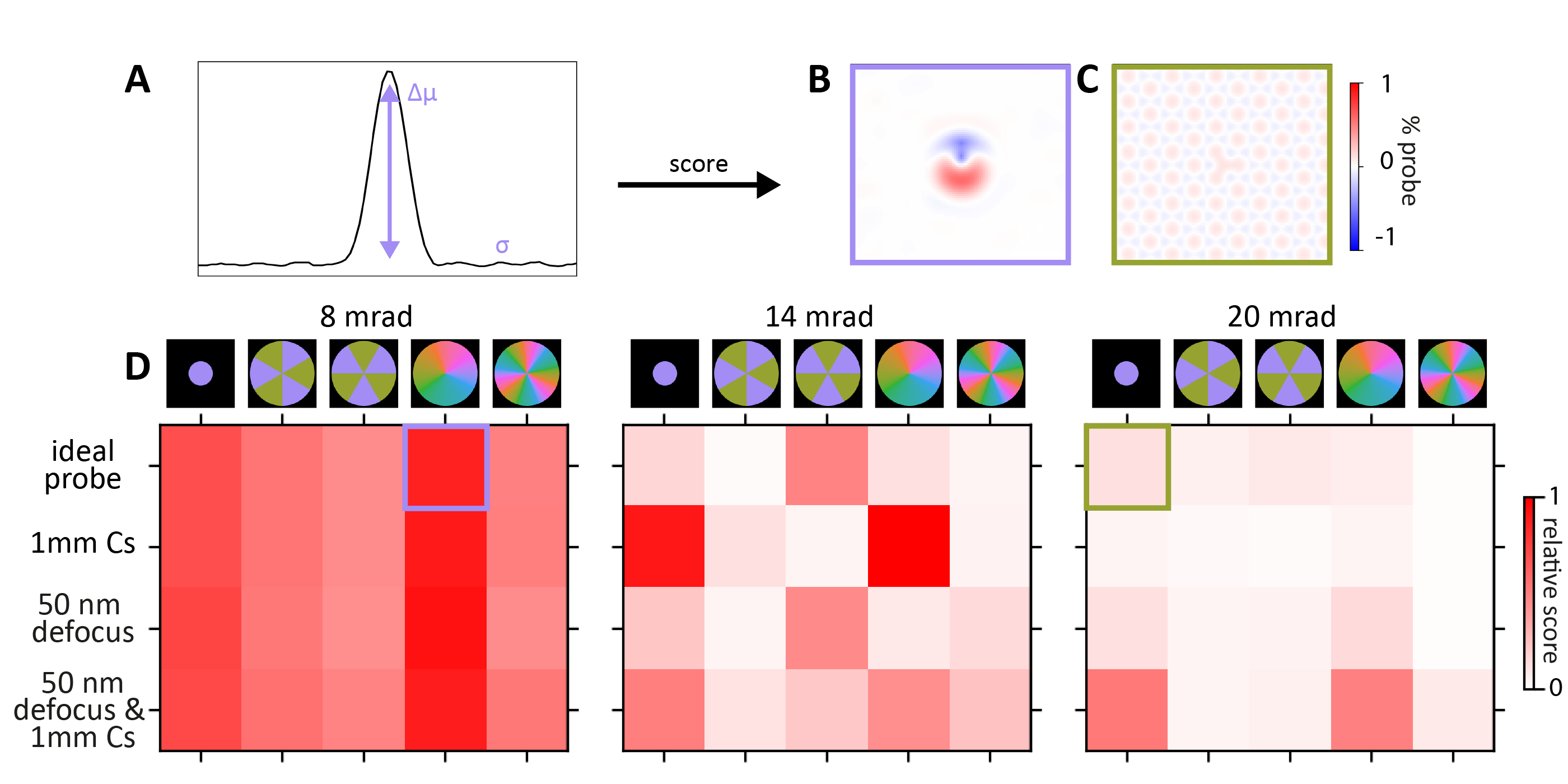}
\caption{(A) The scoring metric evaluates the signal to noise of the defect, which gives high marks to (B) with the defect highlighted above the atomic signal as compared to (C) where we see atomic contrast. (D) Square root of SNR: testing across a range of experimental parameters and detectors suggests small convergence angle and a rotationally varying detector is the best way to highlight the defect.}

\label{fig:score}
\end{figure*}

Fig.\ref{fig:detectors} proposes a number of detectors that can be used to highlight structural information from graphene, but ultimately the image contrast will change with experimental conditions. To determine how robust the contrast is with these detectors, we scored our images based on the signal from the defect versus the background signal, using masks defined in Fig.\ref{fig:SI_mask}.

\begin{equation}
\textrm{SNR} = \frac{\mu_{|defect|}-\mu_{|lattice|}}{\sigma_{|lattice|}}
\label{eq:1}
\end{equation}

To help make the trends more apparent to the reader, we plot the square root of SNR here, with more red squares scoring higher. This scheme gives high marks to images like those in Fig.\ref{fig:score}B, where the defect is highlighted above atomic signals. It also gives low scores to images like those in Fig.\ref{fig:detectors}C, where contributions from the atomic lattice dominate the contrast. This is a beneficial visualization technique in a system like graphene where the lattice is well defined, and one would be interested in quickly and efficiently finding breaks in symmetry. 

By testing the detectors described in Fig.\ref{fig:detectors} against a variety of experimental parameters, we observe that the CoM type detector with a relative small convergence angle (8 mrad) is the most consistently efficient way to highlight a defect (Fig.\ref{fig:score}D). This can be explained by the fact that with a small convergence angle, or large probe in real space, the 4D-STEM experiment is less sensitive to variations in atomic contrast, making other signals more apparent. This result is also the most robust against commonly encountered aberrations, such as spherical aberrations and defocus. 

For atomic resolution images, such as those in Fig.\ref{fig:detectors}, aberrations and defocus will interfere with the ability to resolve atomic information. Although these results are obtained for graphene, this suggests that this rotationally varying detectors and a small convergence angle can be used to highlight defects that break long-range order in other systems as well. 

\subsection*{Amplifying signal with beam modification}
\begin{figure}
\includegraphics[width=\linewidth]{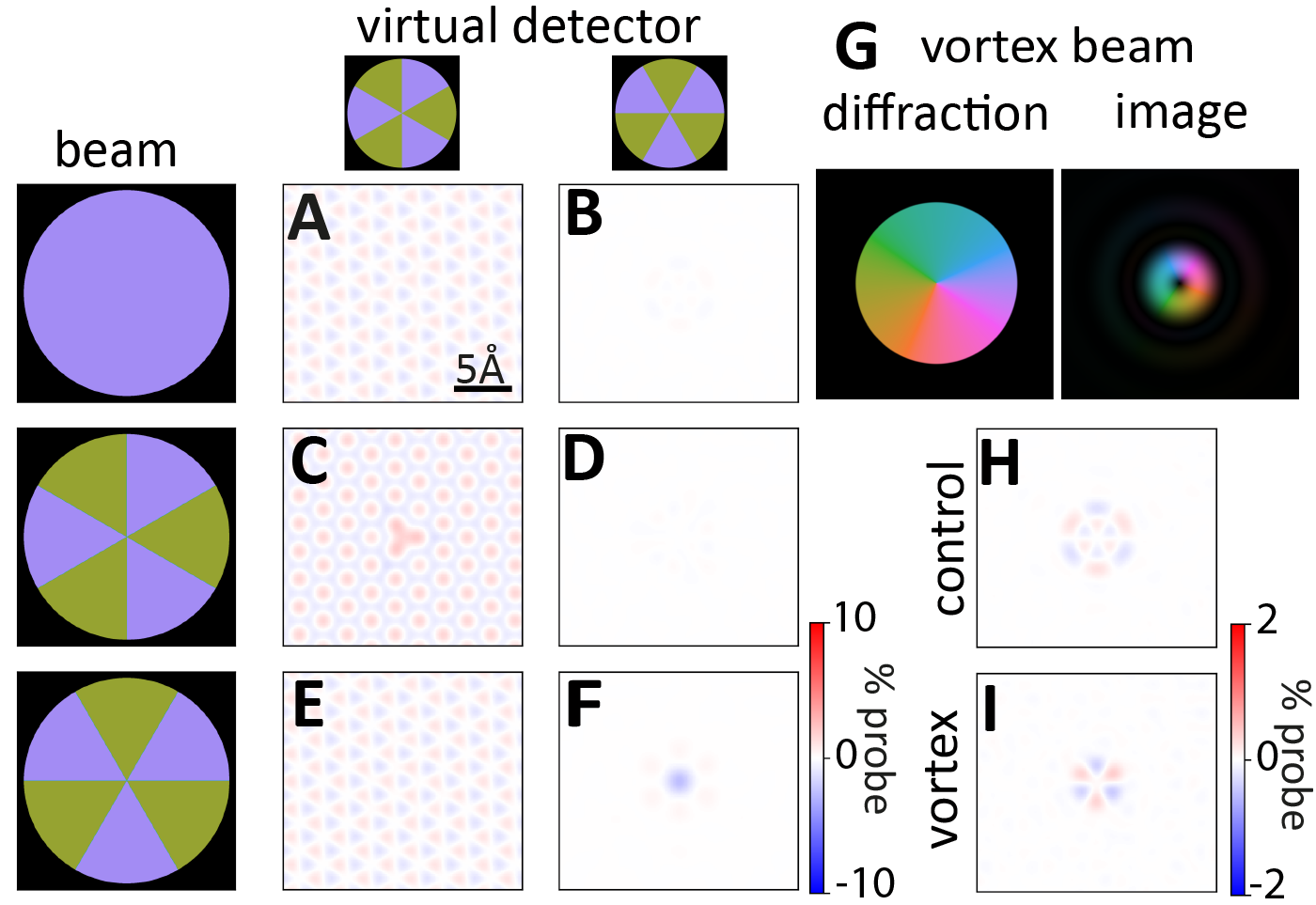}
\caption{With an ideal probe, the wedge shape detectors show (A) atomic and (B) defect contrast depending on orientation. However, if the beam is modified before interaction with the sample (C-F) the contrast is stronger. In (C) and (F) the virtual detectors and beam modification are matched. (C) shows holes in graphene rings and the vacancy, while (F) shows only the defect. (G) A vortex beam can also enhance the signal from the defect as shown in (I) as compared to (H).}
\label{fig:mod}
\end{figure}

\begin{figure*}
\includegraphics[width=\linewidth]{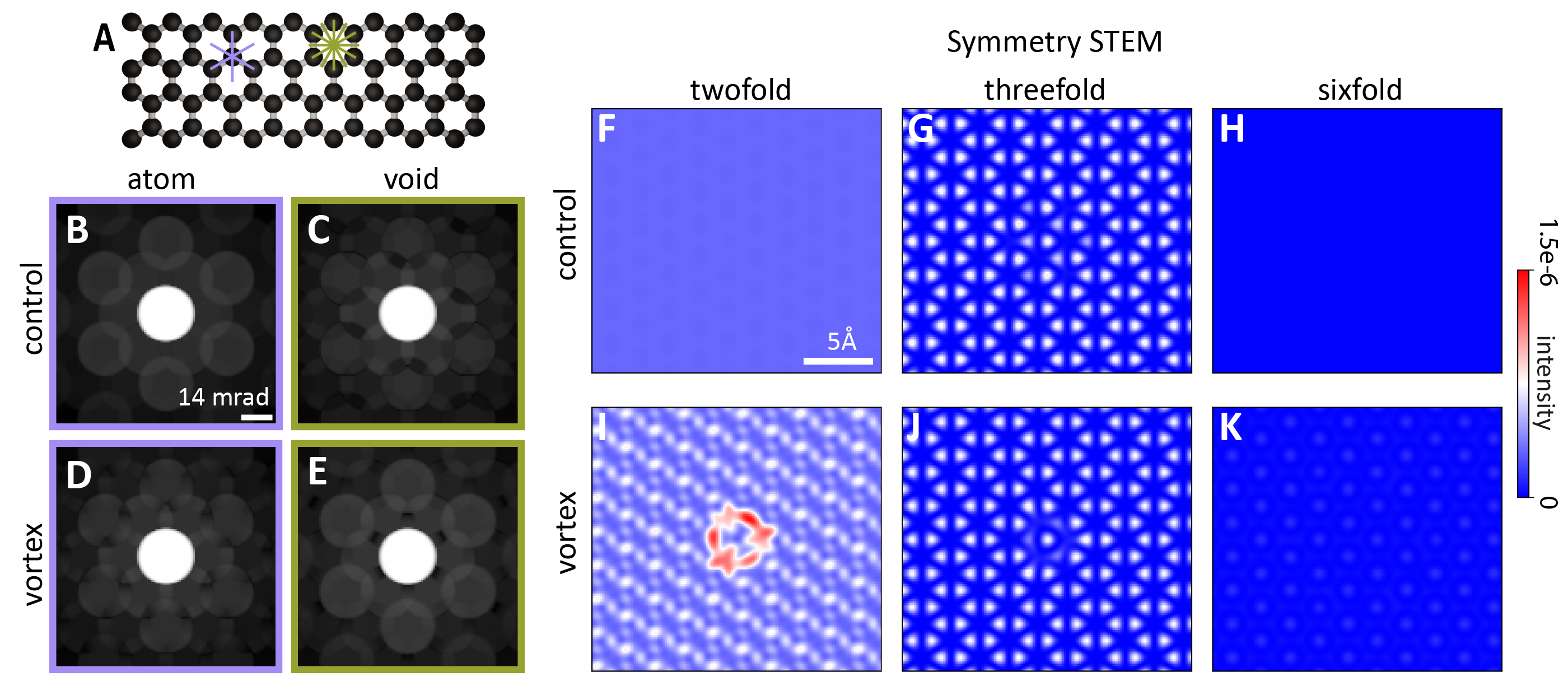}
\caption{(A) Graphene atomic sites and voids have 3- and 6-fold symmetry respectively, which (B-C) appear as 6-fold symmetry in CBED patterns. (D-E) the vortex beam distinguishes between 3- and 6-fold symmetry as the diffraction pattern from the atomic position now has 3- instead of 6-fold symmetry. (F-K) Both probes can be used for finding atomic positions and defects with S-STEM. However, the vortex beam lends itself especially well to S-STEM maps for finding defects, as demonstrated in the 2-fold case (I).}
\label{fig:vortex}
\end{figure*}

Modifying the phase or amplitude of an electron beam before interaction with the sample can also dramatically influence image contrast. In TEM experiments, especially for imaging of weak phase objects such as biological molecules, a Zernike or Volta phase plate is routinely used to enhance image contrast~\citep{WANG2019175, DANEV201787}. In STEM, it has been shown that modification of the probe by inserting a phase plate in the probe forming aperture enhances the signal from both heavy and light elements across a wide range of spatial frequencies ~\citep{ophus2016efficient}.

Combining the concepts of phase plates and wedge detectors, we can evaluate the impact of beam modification on image contrast. In Fig.\ref{fig:mod}A-B, we show the same images as Fig.\ref{fig:detectors}D-E but with a 10x larger contrast range. The signal from the aligned lattice is significantly reduced, and it is no longer easy to see the defect.

In Fig.\ref{fig:mod}C-F, we explore how modifying the incident beam can improve the information transfer from this defect.  On the left the phase and amplitude of the electron beam is plotted, and then to its right are images with two different virtual detectors.  Especially in Fig.\ref{fig:mod}C\&F, we can see how matching detectors and beams can highlight the atomic signal or the defect contrast. The obtained signal from the defect in these images is higher than without beam modification, which leads to a more efficient characterization approach.

Although phase plates provide an exciting avenue for improving image contrast, they are difficult to implement experimentally ~\citep{malac2021phase}. Therefore, we turned to testing this approach with vortex beams, as they have been demonstrated in both TEM and STEM with a variety of beam modification mechanisms ~\citep{uchida2010generation,verbeeck2010production, mcmorran2011electron, verbeeck2012new}. Vortex beams carry orbital angular momentum, and this probe is independent of the material or crystal orientation, allowing for much broader applicability.  The wavefunction of a vortex beam, $\Psi_v(q)$, is defined by:
\begin{equation}
\Psi_v(q) = \Psi(q)e^{im\phi} 
\label{eq:vortex}
\end{equation}
Here,$\Psi(q)$ is the wavefunction of an unmodified beam, m is the quantum number, and $\phi$ is the azimuthal coordinates with respect to the propagation direction of the electron beam. Fig.\ref{fig:mod}G shows a vortex beam in real and reciprocal space with $m=1$.  We observe enhanced contrast from the vortex beam -- for example, comparing Fig.\ref{fig:mod}H\&I, we see stronger intensity and a more localized signal. Fig.\ref{fig:mod}H shows the same control as \ref{fig:mod}B, namely the virtual detectors with a perfect probe. However, here the contrast is scaled to a lower value because the vortex beam is not as efficient as a beam with 3-fold phase contrast.  Although this is not as high transfer of information as with the tailored probe in Fig.\ref{fig:mod}F, the vortex beam provides an improvement in defect detection with reduced beam modification challenges. 

There are other benefits to using an electron vortex beam when studying symmetry, as they are not bound by Friedel's law~\citep{juchtmans2016extension}.  In Fig.\ref{fig:vortex}A, we highlight graphene's symmetry. The atomic sites have 3-fold symmetry, as highlighted with purple bars, while the voids between atoms have 6-fold symmetry, as shown in green. Both of these sites have 6-fold symmetry in reciprocal space, as demonstrated in the diffraction patterns in Fig.\ref{fig:vortex}B and C. We can compare these to CBED patterns taken from the same locations but with a vortex beam (Fig.\ref{fig:vortex}D and E). Here, we see that the atomic site has 3-fold diffraction symmetry, while the void is 6-fold. This experiment highlights the benefits of using the vortex beam to probe symmetry. 

\cite{krajnak2020symmetry} introduced the concept of symmetry STEM, or S-STEM. In this dose-efficient method, the image is based on the cross-correlation between a CBED pattern and the same pattern after a symmetry operation has been applied.  This is a powerful method to understand crystalline order and has been suggested as an effective tool to look for vacancies and other defects.  We explore an extension of this approach to map symmetry using the vortex beam.

The S-STEM maps are effective for highlighting both the graphene structure and the defect. There is a large signal-to-noise difference between the vortex and the control beam for 2-fold symmetry (Fig.\ref{fig:vortex}F\&I). The vortex case produces a stronger signal and highlights the defect. 

We can explore higher symmetries, which will also be different for the conventional and vortex probes. As shown in Fig.\ref{fig:vortex}G and J, the 3-fold S-STEM images show bright contrast at all the atomic sites. The vacancy also has 3-fold symmetry, so it similarly appears to the atomic sites.  With the control beam, we see slightly reduced intensity from the neighboring atoms around the vacancy, as they are no longer in a 3-fold environment. This contrast difference is amplified in the vortex beam case (Fig.\ref{fig:vortex}J). 

The higher 6-fold symmetry map has weak contrast with the control probe. With the vortex probe, we can observe the faint 6-fold spots between atoms.  Fig.\ref{fig:SI_SSTEM} shows these results across a wider range of convergence angles and symmetries. Overall S-STEM provides a powerful tool to map symmetry in 4D-STEM experiments, and a vortex beam could provide a way to make this technique even more interpretable. 

\subsection*{Beyond 2D materials}

\begin{figure*}
\includegraphics[width=\linewidth]{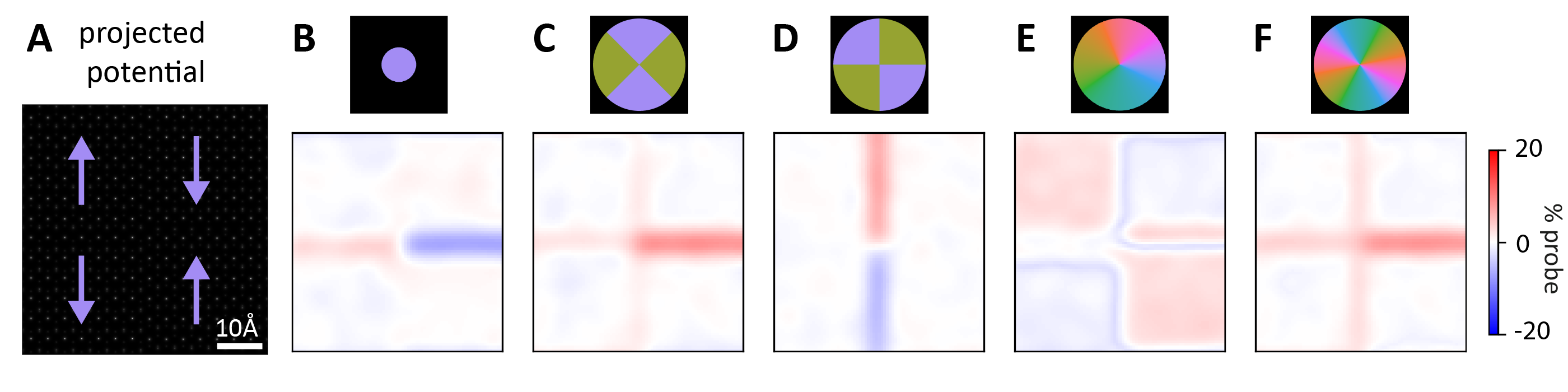}
\caption{(A) PbTiO$_3$ sample with ferroelectric domain boundaries. (B) Bright-field imaging picks up on some walls. (C-D) Wedge-shaped detectors accentuate boundaries. (E) Complex detector highlights the different phases of domains. (F) 2-fold complex detector shows all boundaries.}
\label{fig:pto}
\end{figure*}

Two-dimensional systems, such as graphene, are ideal for testing new virtual detectors and beam modification in 4D-STEM experiments. However, many key materials science challenges are in understanding structures that do not have an inherently two-dimensional architecture. Towards that end, we show simulations of 15 nanometer PbTiO$_3$, a bulk material, which is known to undergo a ferroelectric ordering transition~\citep{shirane1951phase}. We can use the virtual detector design rules established with graphene to see how well they translate to another structure with more complex scattering. Based on our results in Fig. \ref{fig:score}, we use a small convergence angle (2 mrad) and apply similar detectors as shown in Fig.\ref{fig:detectors}.

We explore the PbTiO$_3$ structure with two-dimensional defects, or domain walls \citep{nataf2020domain}, to test the virtual detectors. In our model structure (Fig.\ref{fig:pto}A), the heavy lead atoms are aligned, but the low-Z titanium and oxygen atoms are lined up in opposite directions creating four local domains with two different types of dipoles.  There are three types of domain walls in this structure: head-to-head, tail-to-tail, and opposite orientation. 

The bright field image (\ref{fig:pto}B) identifies the head-to-head and tail-to-tail domain boundaries. These walls have a larger difference in atomic density as compared to the bulk structure, which makes them easier to see with bright field. The opposite orientation domain boundaries are harder to detect.

In Fig.\ref{fig:pto}C, we can see that segmented detectors more efficiently pick up on all the domain boundaries, and Fig.\ref{fig:pto}D shows intensity from the opposite facing domain walls.  The 1-fold rotationally varying detector (Fig.\ref{fig:pto}E) provides domain orientation information, as the upward and downward-facing domains have different contrast. 

The 1-fold center complex detector provides comparable contrast to the magnitude of CoM images. In this sample, the PbTiO$_3$ has minimal y deflection, so the dominant contrast in this image matches the x-component of the CoM images, as shown in Fig.\ref{fig:SI_COM}. The 2-fold rotationally varying detector (Fig.\ref{fig:pto}F) highlights the defects and differentiates between the different kinds of domain boundaries. Overall virtual detectors (Fig.\ref{fig:pto}C-F) provide helpful correlative information to conventional approaches for identifying defects.

One potential challenge in applying virtual detectors in PbTiO$_3$  is the possibility for contrast reversals based on thickness and tilt~\citep{shao2019decoupling, zeltmann2022uncovering}.  To that end, we perform simulations on a variety of thicknesses and mistilts. Fig.\ref{fig:SI_thickness} shows simulations of the same sample but at 1.5 and 2 times thickness (23 and 31 nm respectively). The wedge detectors have the most consistent contrast and pick-up on the boundaries between domains. Fig.\ref{fig:SI_angle} explores how large mistilts of the beam can change the contrast in the image. These simulations are with the same 15nm sample and up to 5mrad of tilting. As with the variations in thickness, there are some contrast reversals, but for the most part these detectors can pick-up on the boundaries. Careful simulations and calibrations would be needed to compare against experiments. However, overall these images suggest that correlative images from different detectors could be helpful for finding defected areas of interest for further study. 

\color{black}

\section{Summary and outlook}

In this work, we explored the use of beam modification and virtual detectors to enhance contrast from defects in 4D-STEM experiments. In systems such as graphene and PbTiO$_3$, where the structure is well characterized, we are most interested in elevating the defect signal above other contrast. We have shown how using a small convergence angle and a rotationally varying detector most efficiently finds these features.  This study focuses on crystalline materials, although in the future we are interested in understanding how these ideas can be extended to less ordered systems. Moreover, the framework used here to evaluate images can be extended in future studies aimed at scoring approaches for materials characterization. We aim to continue exploring the large design space of virtual detectors to highlight material properties.

This study also tests electron beam profiles beyond the conventional probe. There are many practical experimental challenges to phase and amplitude modification of an electron beam~\citep{malac2021phase}, with only further complications in an experiment as proposed in Fig.\ref{fig:mod}, where the orientation of the probe relative to the sample modifies contrast.  We plan continued investigation into phase and amplitude plates, both in experiment and simulation.

\section*{Competing interests}
The authors declare that they have no competing interest.

\section*{Data availability}
Simulation data is available on \href{https://zenodo.org/record/7604087#.Y96W6S-B1mA}{Zenodo}.

\section*{Acknowledgements} \label{sec:acknowledgements}

This material is based upon work supported by the U.S. Department of Energy, Office of Science, Office of Workforce Development for Teachers and Scientists, Office of Science Graduate Student Research (SCGSR) program. The SCGSR program is administered by the Oak Ridge Institute for Science and Education for the DOE under contract number DE‐SC0014664. SMR acknowledges support from the IIN Ryan Fellowship and the 3M Northwestern Graduate Research Fellowship. This material is based upon work supported by the National Science Foundation under Grant No. DMR-1929356. CO acknowledges support from the US Department of Energy Early Career Research Program.  This work made use of the EPIC facility of Northwestern University’s NU\textit{ANCE} Center, which has received support from the SHyNE Resource (NSF ECCS-2025633), the International Institute of Nanotechnology (IIN), and Northwestern's MRSEC program (NSF DMR-1720139).  Work at the Molecular Foundry was supported by the Office of Science, Office of Basic Energy Sciences, of the U.S. Department of Energy under contract number DE-AC02-05CH11231. We thank Steven Zeltmann for helpful feedback on this manuscript.

\newpage

\bibliographystyle{abbrvnat}
\bibliography{main}    





\newpage
\setcounter{section}{0}
\setcounter{page}{1}
\renewcommand\thepage{S.\arabic{page}} 

\setcounter{figure}{0}  
\renewcommand\thefigure{S.\arabic{figure}}

\begin{figure*}[h!]
\includegraphics[width=0.9\linewidth]{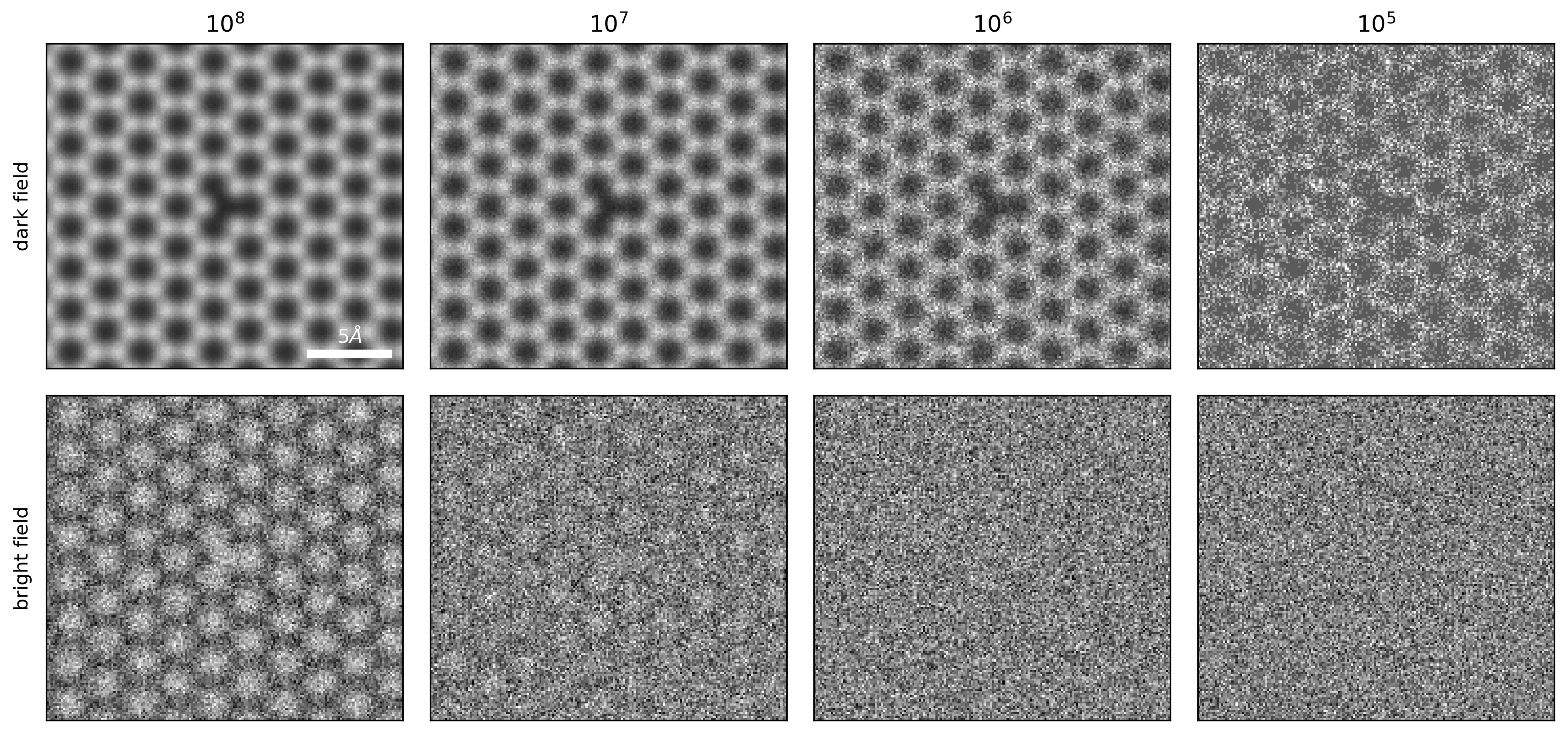}
\caption{These images use a 20mrad convergence angle, and the dose in $e{^-}$\AA $^{-2}$ is listed above. Conventional imaging techniques need high doses to achieve atomic resolution.}
\label{fig:SI_BFDF}
\end{figure*}

\begin{figure*}[h!]
\includegraphics[width=0.9\linewidth]{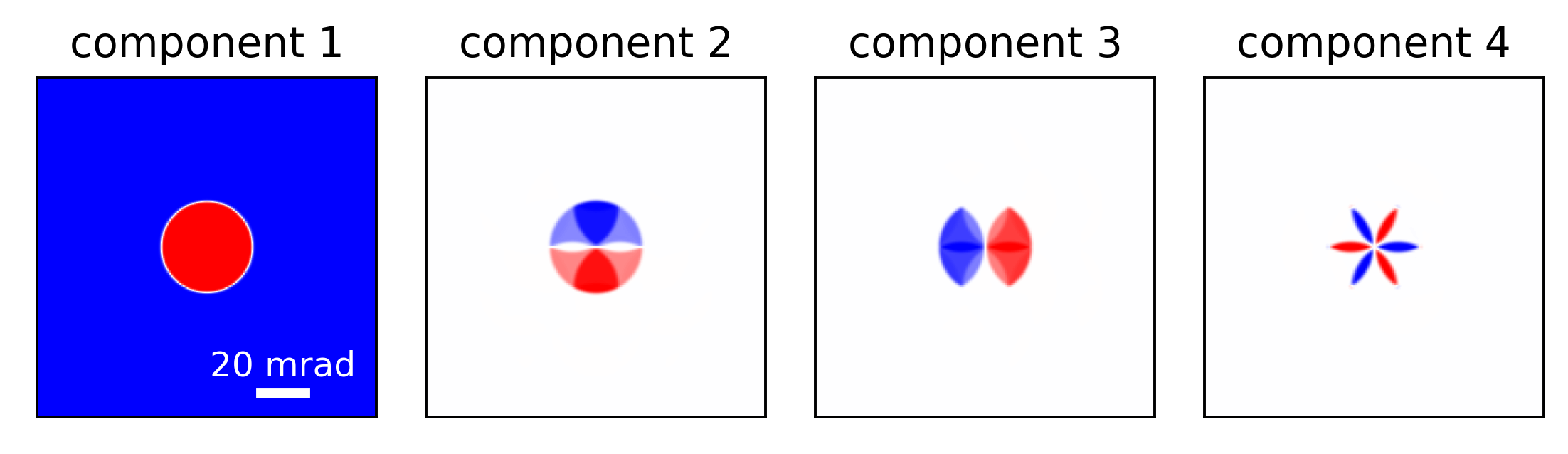}
\caption{First four components of PCA deconvolution of 4D-STEM experiment on pristine graphene.}
\label{fig:SI_PCA}
\end{figure*}

\begin{figure*}[h!]
\includegraphics[width=0.9\linewidth]{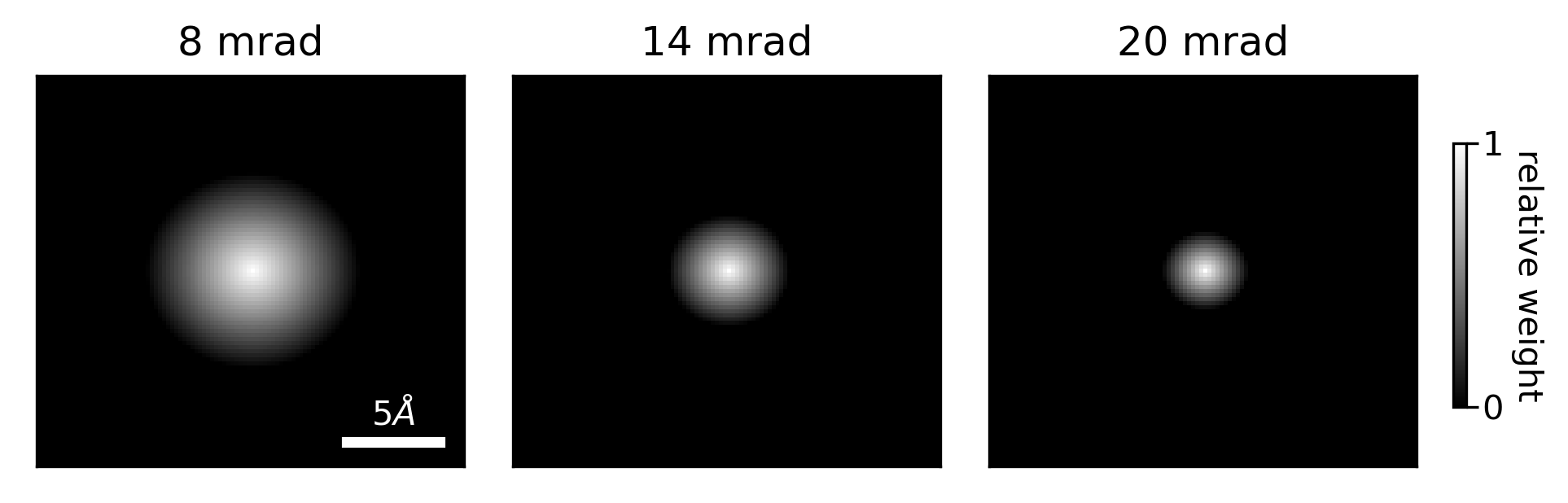}
\caption{Masks used for scoring images in Fig.4. Masks were determined based on the real space probe size. These masks are weighted so that more localized signal close to the defect is counted more strongly.}

\label{fig:SI_mask}
\end{figure*}

\begin{figure*}
\includegraphics[width=0.9\linewidth]{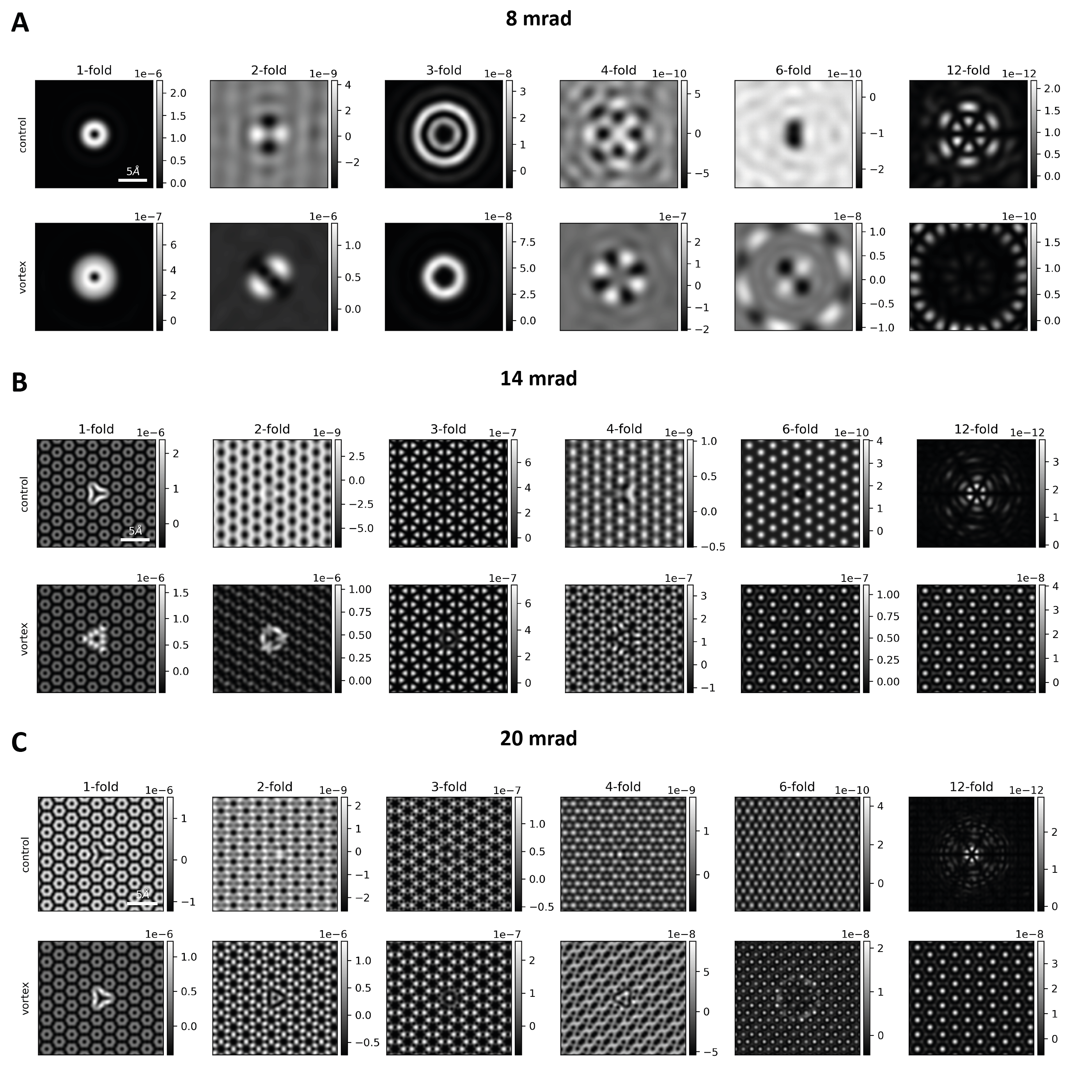}
\caption{Symmetry-STEM reconstructions as a function of convergence angle. }
\label{fig:SI_SSTEM}
\end{figure*}

\begin{figure*}
\centering
\includegraphics[width=0.6\linewidth]{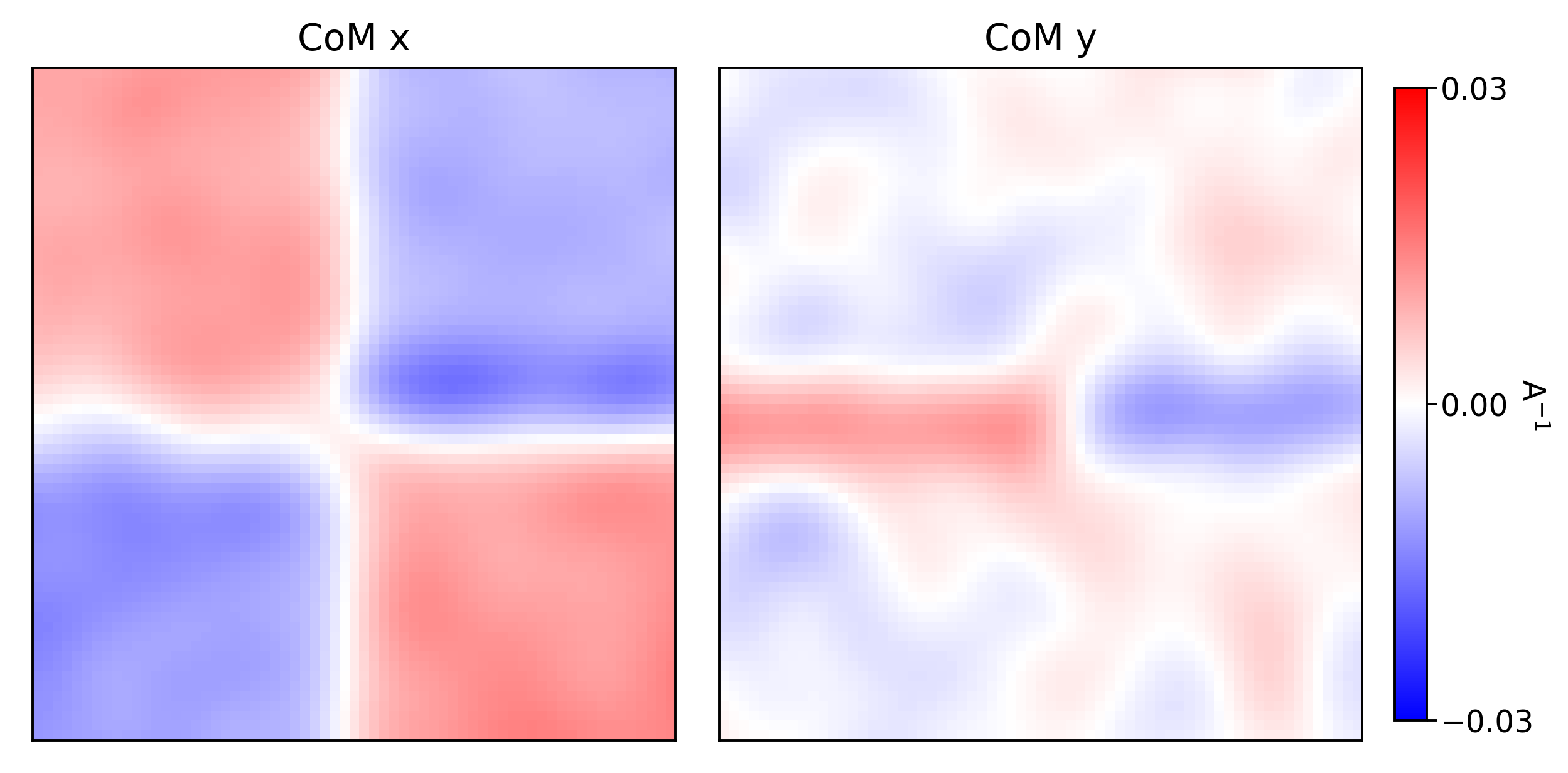}
\caption{PbTiO$_3$ center of mass reconstructions.}
\label{fig:SI_COM}
\end{figure*}

\begin{figure*}
\includegraphics[width=0.9\linewidth]{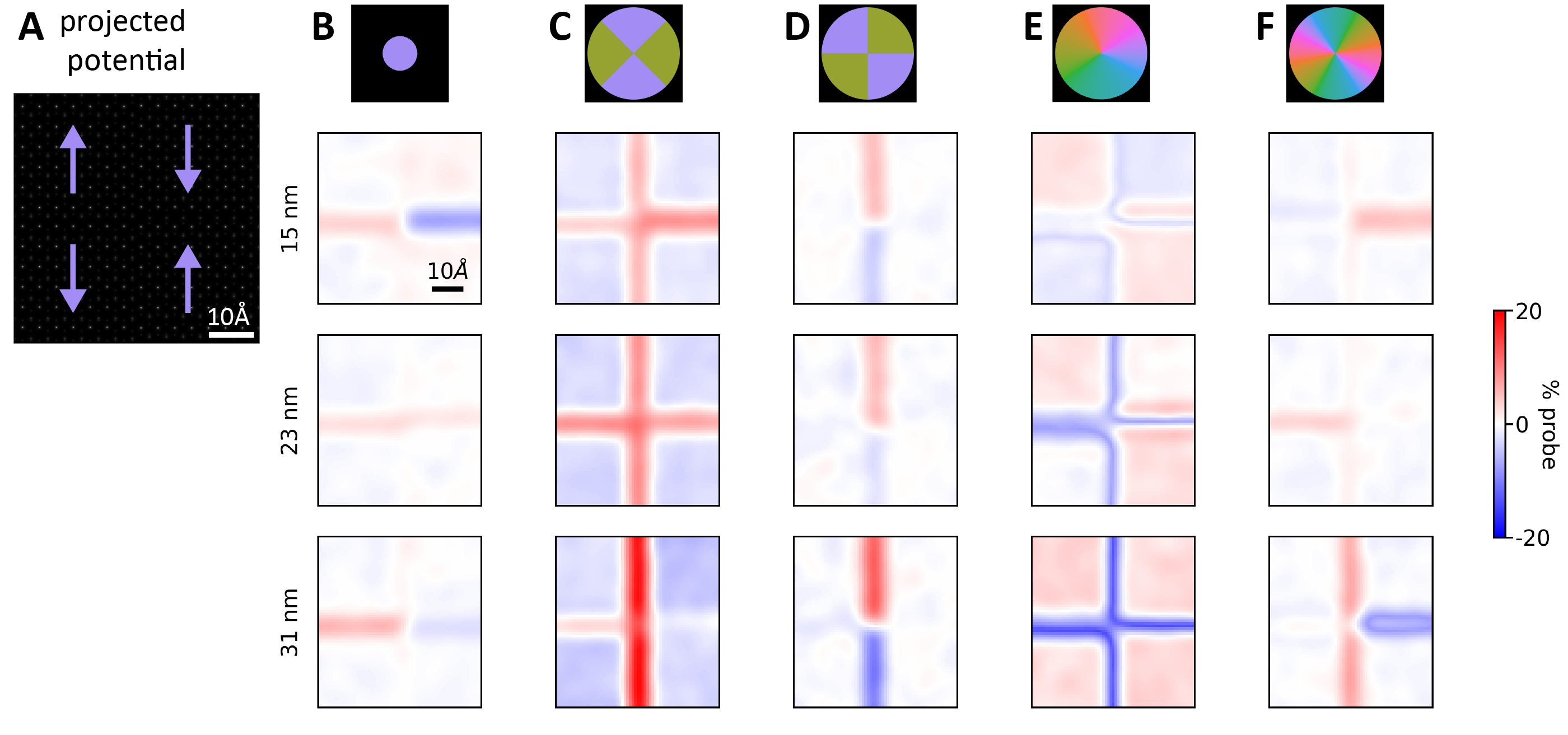}
\caption{Fig.7 reconstructions as a function of thickness.}
\label{fig:SI_thickness}
\end{figure*}

\begin{figure*}
\includegraphics[width=0.9\linewidth]{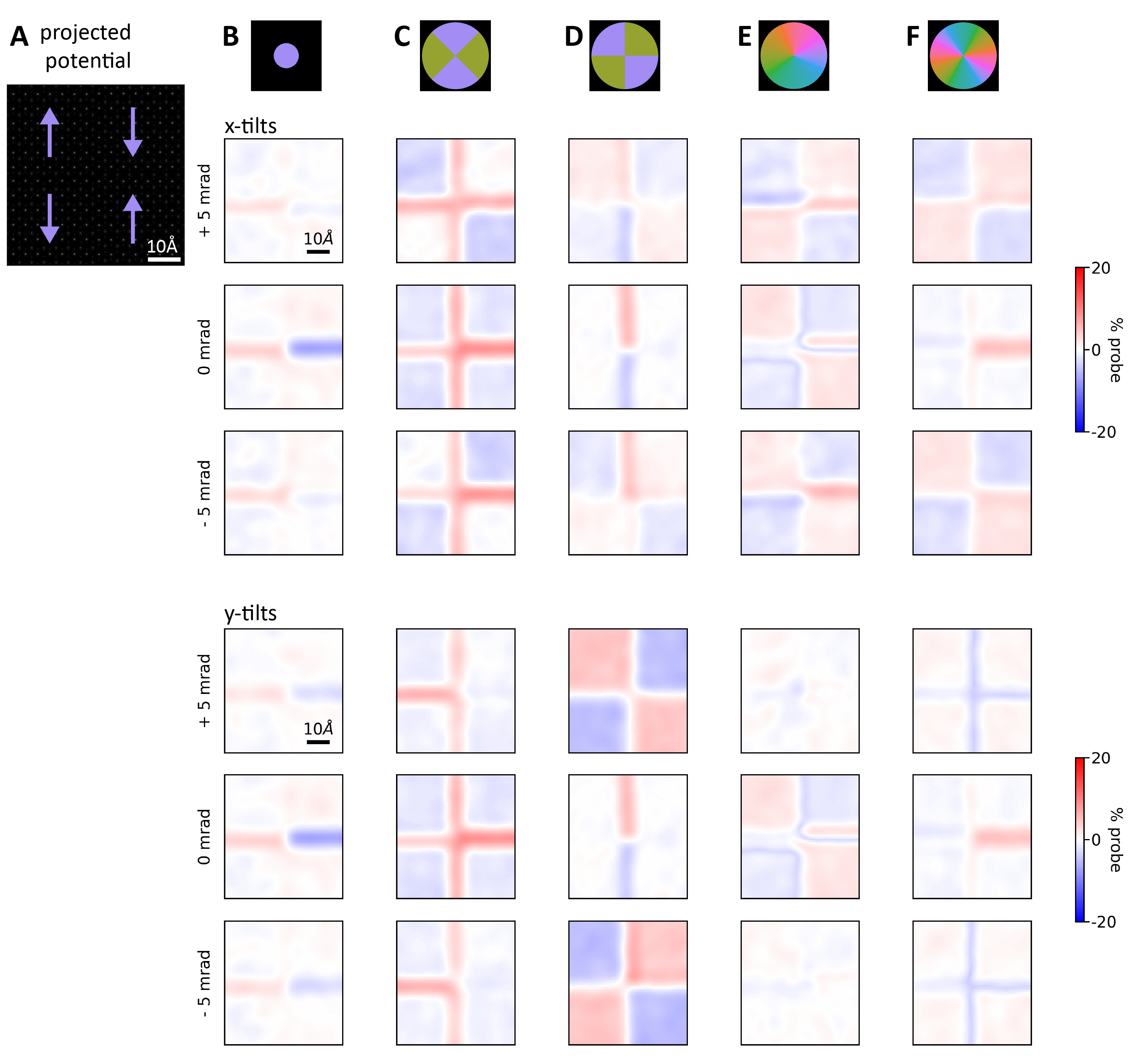}
\caption{Fig.7 reconstructions with small mistilts.}
\label{fig:SI_angle}
\end{figure*}

\end{document}